\documentclass[aps,reprint,pre,superscriptaddress,amsmath,amssymb,longbibliography]{revtex4-1}
\usepackage{graphicx}
\usepackage{amsmath}
\usepackage{bm}
\usepackage{physics}

\usepackage{color}

\definecolor{CLBlue}{rgb}{0, .3, .6}

\begin{document}

\title{Exact minimax entropy models of large--scale neuronal activity}
%\title{Trees of maximally informative correlations in large--scale neuronal activity}

\author{Christopher W.~Lynn}
\email{Corresponding author: christopher.lynn@yale.edu}
\affiliation{Initiative for the Theoretical Sciences, The Graduate Center, City University of New York, New York, NY 10016, USA}
\affiliation{Joseph Henry Laboratories of Physics, Princeton University, Princeton, NJ 08544, USA}
\affiliation{Department of Physics, Quantitative Biology Institute, and Wu Tsai Institute, Yale University, New Haven, CT 06520, USA}
%\affiliation{Quantitative Biology Institute, Yale University, New Haven, CT 06520, USA}
%\affiliation{Wu Tsai Institute, Yale University, New Haven, CT 06520, USA}
\author{Qiwei Yu}
\affiliation{Lewis--Sigler Institute for Integrative Genomics, Princeton University, Princeton, NJ 08544, USA}
\author{Rich Pang}
\affiliation{Princeton Neuroscience Institute, Princeton University, Princeton, NJ 08544, USA}
%\author{Caroline M.~Holmes}
%\affiliation{Joseph Henry Laboratories of Physics, Princeton University, Princeton, NJ 08544, USA}
\author{Stephanie E.~Palmer}
\affiliation{Department of Organismal Biology and Anatomy, University of Chicago, Chicago, IL 60637, USA}
\affiliation{Department of Physics, University of Chicago, Chicago, IL 60637, USA}
\author{William Bialek}
\affiliation{Joseph Henry Laboratories of Physics, Princeton University, Princeton, NJ 08544, USA}
\affiliation{Lewis--Sigler Institute for Integrative Genomics, Princeton University, Princeton, NJ 08544, USA}
\affiliation{Center for Studies in Physics and Biology, Rockefeller University, New York, NY 10065 USA}

\date{\today}

\begin{abstract}
In the brain, fine--scale correlations combine to produce macroscopic patterns of activity. However, as experiments record from larger and larger populations, we approach a fundamental bottleneck: the number of correlations one would like to include in a model grows larger than the available data. In this undersampled regime, one must focus on a sparse subset of correlations; the optimal choice contains the maximum information about patterns of activity or, equivalently, minimizes the entropy of the inferred maximum entropy model. Applying this ``minimax entropy" principle is generally intractable, but here we present an exact and scalable solution for pairwise correlations that combine to form a tree (a network without loops). Applying our method to over one thousand neurons in the mouse hippocampus, we find that the optimal tree of correlations reduces our uncertainty about the population activity by 14\% (over 50 times more than a random tree). Despite containing only 0.1\% of all pairwise correlations, this minimax entropy model accurately predicts the observed large--scale synchrony in neural activity and becomes even more accurate as the population grows. The inferred Ising model is almost entirely ferromagnetic (with positive interactions) and exhibits signatures of thermodynamic criticality. These results suggest that a sparse backbone of excitatory interactions may play an important role in driving collective neuronal activity.
\end{abstract}

%\pacs{asdfasdf}
%\keywords{keywords}

\maketitle


\section{Introduction}
\label{sec_intro}

Understanding how collective behaviors emerge from webs of fine--scale interactions is a central goal in statistical mechanics approaches to networks of neurons \cite{wiener_58, cooper_73, little_74, hopfield1982,amit_89, hertz+al_91}. At the same time, exploration of the brain has been revolutionized by experimental methods that monitor, simultaneously, the electrical activity of hundreds or even thousands of neurons \cite{segev+al2004, litke+al2004, chung2019high, dombeck2010functional, tian2012neural, demas+al2021, steinmetz2021neuropixels}. One approach to connecting these new data with statistical physics models is maximum entropy, in which we construct the maximally disordered model that is consistent with measured expectation values \cite{Jaynes-01}. In particular, is seems natural to build models that match the mean activity of individual neurons and the correlations between pairs of neurons.  These pairwise maximum entropy models have been strikingly successful in describing collective behavior not only in networks of real neurons, but also in the evolution of protein families, the dynamics of genetic networks, flocks of birds, and social networks \cite{Schneidman-01, Nguyen-01, Meshulam-17, Meshulam-02, Tkacik-03, Lezon-01, weigt+al_09, marks2011protein, lapedes2012using, Bialek-01, russ2020evolution, Lynn-04}.
%Information about the interactions between neurons is encoded in their correlations, which can now be measured %experimentally among thousands of cells 
%To predict the macroscopic activity patterns that arise from a set of fine--scale correlations, one must consider the hypothetical system that is consistent with the given statistics but is otherwise maximally disordered \cite{shannon1948mathematical, Jaynes-01}. Such maximum entropy models have provided key insights into the emergence of large--scale order in living systems, from populations of neurons and whole--brain networks to flocks of birds, genetic interactions, and human social networks \cite{Schneidman-01, Nguyen-01, Meshulam-17, Meshulam-02, Tkacik-03, Lezon-01, weigt+al_09, marks2011protein, lapedes2012using, Bialek-01, russ2020evolution, Lynn-04}. 

But as experiments progress to record from larger and larger numbers of neurons, we face a combinatorial explosion. Even if we focus on pairwise correlations,
%, the number of possible correlations that one could include in a model grows exponentially. Even if we focus on the simplest statistics---those involving pairs of variables---
the number of correlations approaches the number of independent samples in modern experiments \cite{segev+al2004, litke+al2004, chung2019high, dombeck2010functional, tian2012neural, demas+al2021, steinmetz2021neuropixels}. In this undersampled regime, one is forced to select only a sparse subset of the correlations to include in any model. While constructing an accurate model with only a small number of correlations may seem hopeless, one can draw inspiration from statistical physics, where effective descriptions of macroscopic phenomena can often ignore many of the microscopic details.

Here, given restrictions on the number and structure of correlations we can include in a model, we seek to identify the ones that contain the maximum information about system activity. We demonstrate that the optimal correlations are those that induce the maximum entropy model with minimum entropy \cite{Zhu-01}. %As we will see, s
Solving this minimax entropy problem is generally infeasible. But for pairwise correlations that form a tree (a network without loops), the entropy reduction decomposes into a sum over connected pairs; the advantages of tree structure in models of neural activity have been appreciated in other contexts \cite{Prentice-01}.
%amount of information contained in each correlation reduces to the mutual information between the two elements. This observation 
This decomposition reduces the minimax entropy problem to a minimum spanning tree problem, which can be solved exactly and efficiently \cite{Chow-01, Nguyen-01}. The result is a framework for uncovering the maximally informative tree of correlations in very large systems \cite{Lynn-15}.

We apply our method to investigate the collective activity of $N\sim1500$ neurons in the mouse hippocampus \cite{Gauthier-01}. While most pairs of neurons are only weakly correlated, some rare pairs have mutual information orders of magnitude larger than average. By focusing on these exceptionally strong correlations, our minimax entropy model captures 50 times more information than a random tree and, despite containing only 0.1\% of all pairwise correlations, produces realistic large--scale synchrony in activity. Moreover, the model becomes even more accurate as the population grows, providing hope for statistical physics descriptions of even larger systems.

The paper is organized as follows. In \S\ref{sec_minimax}, we define the minimax entropy problem and present a solution for trees of pairwise correlations. In \S\ref{sec_data}, we review a relatively recent experiment on large--scale recordings of neuronal activity in the mouse hippocampus. In \S\ref{sec_synchrony}, we demonstrate that the optimal tree of correlations produces realistic patterns of synchronized activity. In \S\ref{sec_model}, we investigate the structural properties of the optimal tree and the functional properties of the induced Ising model. In \S\ref{sec_scaling}, we show that the minimax entropy model becomes more accurate for larger populations, and then in \S\ref{sec_thermo} we investigate the thermodynamic properties of the minimax entropy model, finding that the real system is poised at a special point in its phase diagram. %discovering signatures of critical behavior. 
Finally, in \S\ref{sec_conclusions} we provide conclusions and outlook.

\section{Minimax entropy trees}
\label{sec_minimax}

\subsection{Maximum entropy models}

Consider a system of $N$ elements $i = 1,\hdots,N$ with states $\bm{x} = \{x_i\}$, where $x_i$ is the state of element $i$. From experiments, we have access to $M$ samples of the system activity $\bm{x}^{(m)}$, where $m = 1,\hdots, M$. Our knowledge about the system is defined by observables, which can be represented as expectation values
\begin{equation}
\left<f(\bm{x})\right>_\text{exp} = \frac{1}{M}\sum_{m = 1}^M f(\bm{x}^{(m)}),
\end{equation}
where $f(\bm{x})$ is an arbitrary function of the state $\bm{x}$. For example, one could measure the average states of individual elements $\left<x_i\right>_\text{exp}$ or the correlations among multiple elements $\left<x_ix_j\right>_\text{exp}$, $\left<x_ix_jx_k\right>_\text{exp}$, and so on. Given a set of $K$ observables $\mathcal{O} = \{f_\nu(\bm{x})\}$, where $\nu = 1,\hdots, K$, the most unbiased prediction for the distribution over states is the maximum entropy model \cite{shannon1948mathematical, Jaynes-01}
\begin{equation}
\label{eq_P}
P_\mathcal{O}(\bm{x}) = \frac{1}{Z}\text{exp}\Bigg[-\sum_{\nu = 1}^K \lambda_\nu f_\nu(\bm{x})\Bigg],
\end{equation}
where $Z$ is the normalizing partition function, and the parameters $\lambda_\nu$ ensure that the model matches the experimental observations, such that
\begin{equation}
\label{eq_obs}
\left<f_\nu(\bm{x})\right> = \left<f_\nu(\bm{x})\right>_\text{exp}.
\end{equation}

To have control over errors in the $K$ expectation values, we must have $K \ll MN$. But as experiments record from larger systems, one is confronted with an explosion of possible observables. The total number of correlations grows exponentially with $N$, and even the $K\propto N^2$ pairwise correlations violate the good sampling condition as $N$ grows large. Thus, to avoid sampling problems, one must focus on a sparse subset of correlations. Here we arrive at the central question: Among a large set of observables, which should we choose to include in a model?

\subsection{Minimax entropy principle}

Suppose we want to find the set of observables $\mathcal{O}$ that yields the most accurate description of the system. We can choose $\mathcal{O}$ to maximize the log--likelihood of the model $P_\mathcal{O}$ or, equivalently, minimize the KL divergence with respect to the data $D_\text{KL}(P_\text{exp} || P_\mathcal{O})$. Due to the form of $P_\mathcal{O}$ in Eq.~(\ref{eq_P}), the KL divergence simplifies to a difference in entropies
\begin{align}
D_\text{KL}(P_\text{exp} ||P_\mathcal{O}) &= \left<\log \frac{P_\text{exp}(\bm{x})}{P_\mathcal{O}(\bm{x})}\right>_\text{exp} \nonumber \\
&= \log Z + \frac{1}{\ln 2}\sum_\nu \lambda_\nu \left<f_\nu(\bm{x})\right>_\text{exp} - S_\text{exp} \nonumber \\
&= S_\mathcal{O} - S_\text{exp},
\end{align}
where the final equality follows from Eq.~(\ref{eq_obs}), and entropies $S$ are measured in bits. We therefore find that the optimal observables $\mathcal{O}$ are the ones that minimize the entropy of the maximum entropy model $S_\mathcal{O}$. This is the ``minimax entropy" principle, which was proposed 25 years ago but has received relatively little attention \cite{Zhu-01}.

In addition to providing the best description of the data, the optimal observables $\mathcal{O}$ can also be viewed as containing the maximum information about the system. If we begin by observing each element individually, then we only have access to the marginal distributions $P_i(x_i)$; in this case, the maximum entropy model is the independent distribution $P_\text{ind}(\bm{x}) = \prod_i P_i(x_i)$ with entropy $S_\text{ind}$. If, in addition to the marginals, we also observe some of the correlations between elements \cite{Schneidman-02}, this knowledge reduces our uncertainty about the system by an amount $I_\mathcal{O} = S_\text{ind} - S_\mathcal{O} \ge 0$. Thus, by minimizing $S_\mathcal{O}$, the optimal observables $\mathcal{O}$ also maximize the information $I_\mathcal{O}$ contained in the observed correlations.

In practice, applying the minimax entropy principle poses two distinct challenges. First, for each set of observables $\mathcal{O} = \{f_\nu\}$, one must solve the traditional maximum entropy problem; that is, one must compute the parameters $\lambda_\nu$ such that the model $P_\mathcal{O}$ matches the expectations $\left<f_\nu(\bm{x})\right>_\text{exp}$ in the data. Second, one must repeat this calculation for all sets of observables $\mathcal{O}$ to find the one that minimizes the entropy $S_\mathcal{O}$. This search process is generally intractable. In what follows, we study a class of observables that admits an exact and efficient solution, enabling statistical physics models of very large systems.

\subsection{Trees of pairwise correlations}

For simplicity, we focus on binary variables $x_i = 0,1$, for which the marginals $P_i(x_i)$ are defined by the averages $\left<x_i\right>$. In the search for sources of order in a system, one might begin with the simplest correlations: those between pairs of elements $\left<x_ix_j\right>$. In populations of $N \sim 100$ neurons, one often has sufficient data to fit all the pairwise correlations, which can be very effective in capturing key features of the collective activity \cite{Schneidman-01, Meshulam-17, Meshulam-02}. But this corresponds to $K\propto N^2$ constraints, and at large $N$ we will violate the good sampling condition $K \ll NM$. To avoid undersampling, we are forced to select a sparse subset of pairwise correlations, which can be visualized as a network $\mathcal{G}$ with edges defining the observed correlations between variables. Each network induces a maximum entropy model
\begin{equation}
\label{eq_PG}
P_\mathcal{G}(\bm{x}) = \frac{1}{Z}\text{exp}\Bigg[\sum_{(ij)\in \mathcal{G}} J_{ij}x_ix_j + \sum_i h_ix_i\Bigg],
\end{equation}
where the parameters $h_i$ and $J_{ij}$ enforce the constraints on $\left<x_i\right>$ and $\left<x_ix_j\right>$ in $\mathcal{G}$, respectively. The minimax entropy principle tells us that we should find the network $\mathcal{G}$ (within some allowed set) that produces the maximum entropy model $P_\mathcal{G}$ with minimum entropy $S_\mathcal{G}$.

In statistical physics, calculations are difficult in part due to feedback loops. By eliminating loops, many statistical physics models become tractable, as in one--dimensional systems or on Bethe lattices \cite{sethna_06}. In the Ising model---which is equivalent to Eq.~(\ref{eq_PG})---if the interactions $J_{ij}$ lie on a tree $\mathcal{T}$ (or a network without loops), then one can efficiently compute the partition function $Z$ and all statistics of interest (see Appendix \ref{app_Ising}). Inverting this procedure, one can begin with the averages $\left<x_i\right>$ and the correlations $\left<x_ix_j\right>$ on a tree $\mathcal{T}$ and analytically derive the maximum entropy parameters \cite{Chow-01, Nguyen-01}:
\begin{align}
\label{eq_J}
J_{ij} &= \ln \left[\frac{\left<x_ix_j\right>\left(1 - \left<x_i\right> - \left<x_j\right> + \left<x_ix_j\right>\right)}{\left(\left<x_i\right> - \left<x_ix_j\right>\right)\left(\left<x_j\right> - \left<x_ix_j\right>\right)}\right], \\
\label{eq_h}
h_i &= \ln \frac{\left<x_i\right>}{1 - \left<x_i\right>} \\
& \quad\quad + \sum_{j \in \mathcal{N}_i} \ln \left[\frac{\left(1 - \left<x_i\right>\right)\left(\left<x_i\right> - \left<x_ix_j\right>\right)}{\left<x_i\right>\left(1 - \left<x_i\right> - \left<x_j\right> + \left<x_ix_j\right>\right)}\right], \nonumber
\end{align}
where $\mathcal{N}_i$ represents the neighbors of $i$ in $\mathcal{T}$ (see Appendix \ref{app_maxent}). Since each tree contains $N-1$ correlations, the total number of observables is $K = 2N-1$, and so we are well sampled if the number of independent samples obeys $M \gg 2$.

Equations (\ref{eq_J}) and (\ref{eq_h}) solve the maximum entropy problem for the distribution $P_\mathcal{T}$, but we still need to search over all of the $N^{N-2}$ trees to find the one that minimizes the entropy $S_\mathcal{T}$. This search simplifies significantly by noticing that the information $I_\mathcal{T}$ decomposes into a sum over the connections $(ij)$ in $\mathcal{T}$,
\begin{equation}
\label{eq_I}
I_\mathcal{T} = S_\text{ind} - S_\mathcal{T} = \sum_{(ij)\in\mathcal{T}} I_{ij},
\end{equation}
where $I_{ij}$ is the mutual information between $i$ and $j$ (see Appendix \ref{app_info}) \cite{Chow-01, Nguyen-01}. Note that for pairs $(ij)\in\mathcal{T}$, the mutual information $I_{ij}$ is the same in the model and the data, so we can compute the entropy $S_\mathcal{T}$ directly from the data without constructing the model itself.

\begin{figure}[t]
\centering
\includegraphics[width = \columnwidth]{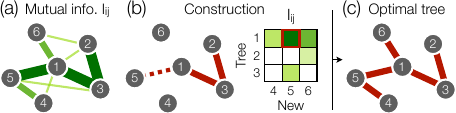} \\
\caption{Constructing the optimal tree. (a) Visualization of the mutual information $I_{ij}$ (edges) between elements in a system (nodes), with darker, thicker edges reflecting larger $I_{ij}$. (b) Illustration of Prim's algorithm. At each step, we consider the mutual information $I_{ij}$ between elements in the tree and those not yet connected (matrix). We then connect the two elements with the largest $I_{ij}$ (dashed edge) and repeat until all elements have been added. (c) Optimal tree that maximizes minimizes $S_\mathcal{T}$ and maximizes $I_\mathcal{T}$. \label{fig_minimax}}
\end{figure}

Equation (\ref{eq_I}) tells us that the tree with the minimum entropy $S_\mathcal{T}$ is the one with the largest total mutual information. Identifying this optimal tree is a minimum spanning tree problem \cite{Nguyen-01}, which can be solved efficiently using a number of different algorithms \cite{Moore-01}. To begin, one computes the mutual information $I_{ij}$ between all elements [Fig.~\ref{fig_minimax}(a)]. One can then grow the optimal tree by greedily connecting the element $i$ in the tree to the new element $j$ with the largest mutual information $I_{ij}$ [Fig.~\ref{fig_minimax}(b)]; this is Prim's algorithm, which runs in $O(N^2)$ time [Fig.~\ref{fig_minimax}(c)]. Thus, by restricting to trees of pairwise correlations, we can solve the minimax entropy problem exactly, even at very large $N$.

\section{Large--scale neuronal activity}
\label{sec_data}

We ultimately seek to explain the collective behaviors of very large networks.
%key features of living systems. 
However, each tree only contains a vanishingly small fraction $2/N$ of all pairwise correlations; and even if we have access to all of the pairwise statistics, there's still no guarantee of success. Can such a sparse set of observations capture something important about the system as a whole?

\begin{figure}[t]
\centering
\includegraphics[width = \columnwidth]{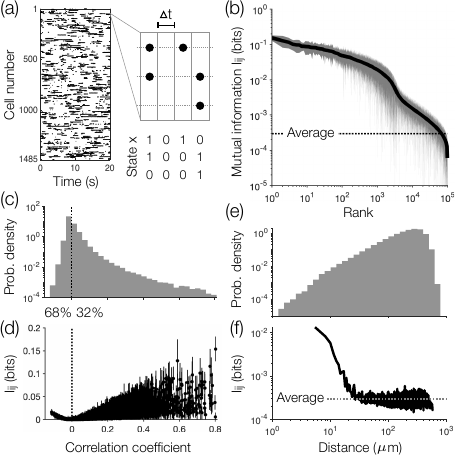} \\
\caption{Collective activity in a large population of neurons. (a) Time series of neuronal activity in the mouse hippocampus, where each dot represents an active neuron (see \cite{Gauthier-01} for experimental details). States $\bm{x} = \{x_i\}$ represent the population activity within a single window of width $\Delta t = 1/30\,\text{s}$. (b) Ranked order of significant mutual information $I_{ij}$ in the population. Solid line and shaded region reflect estimates and errors (two standard deviations) after correcting for finite data (see Appendix \ref{app_MI}). (c) Distribution of correlation coefficients over neuron pairs, with percentages indicating the fraction of positively and negatively correlated pairs. (d) Mutual information $I_{ij}$ versus correlation coefficient, where each point represents a distinct neuron pair. Estimates and errors are the same in (b). (e) Distribution of physical distances between neurons. (f) Average mutual information $I_{ij}$ as a function of physical distance, computed in bins that contain $500$ pairs each; note that individual pairs vary widely around this average.
\label{fig_data}}
\end{figure}

To answer this question, we consider patterns of electrical  activity in $N=1485$ neurons in the hippocampus of a mouse, recorded in a recent experiment \cite{Gauthier-01}. Mice are genetically engineered so that their neurons express a protein whose fluorescence is modulated by calcium concentration, which in turn follows the electrical activity of the cells. This fluorescence is recorded using a scanning two--photon microscope as the mouse runs in a virtual environment. The signal from each cell consists of a quiet background punctuated by short bursts of activity \cite{Meshulam-17}, providing a natural binarization into active ($x_i = 1$) or silent ($x_i = 0$) within each video frame [Fig.~\ref{fig_data}(a)]. Capturing images at $30\,\text{Hz}$ for $39\,\text{min}$ yields $M\sim 7\times 10^4$ samples of the collective state $\bm{x} = \{x_i\}$, but these are not all independent. Nonetheless, we can still estimate the mutual information $I_{ij}$ with small errors after correcting for finite data effects (see Appendix \ref{app_MI}).

Among all $\sim 10^6$ pairs of neurons, only $9\%$ exhibit significant mutual information with values shown in Fig.~\ref{fig_data}(b). We see that a small number of pairs contain orders of magnitude more information than average ($\bar{I} = 2.9\times 10^{-4}$ bits). This heavy--tailed distribution provides hope for a tree of correlations that contains much more information than typical $I_\mathcal{T} \gg (N-1)\bar{I}$. Additionally, while most pairs of cells are negatively correlated [Fig.~\ref{fig_data}(c)], the strongest mutual information corresponds to positive correlations [Fig.~\ref{fig_data}(d)]. And while most neurons are far from one another [Fig.~\ref{fig_data}(e)], larger values of $I_{ij}$ are concentrated among pairs of cells that are close to one another, as can be seen by plotting the mean mutual information as a function of distance
%many of the strong mutual information are concentrated among neighboring cells 
[Fig.~\ref{fig_data}(f)]. Together, these observations suggest that a backbone of positively correlated and physically proximate neurons may provide a large amount of information about the collective neural activity.

\section{Predictions of minimax entropy model}
\label{sec_synchrony}

Constructing the minimax entropy tree (Fig.~\ref{fig_minimax}), we find that that it captures $I_\mathcal{T} = 26.2$ bits of information. This reduces our uncertainty about the population activity by $I_\mathcal{T}/S_\text{ind} = 14.4\%$, which is equivalent to freezing the states of 214 randomly selected neurons. For comparison, we consider two additional networks: (i) a random tree, which represents a typical collection of correlations, and (ii) the tree of minimum physical distances, which reflects the fact that neighboring neurons are more likely to be strongly correlated [Fig.~\ref{fig_data}(f)]. The optimal tree captures over twice as much information as the minimum distance tree and over 50 times more than random.
%, thus establishing a sparse network of highly informative correlations.

\begin{figure}[b]
\centering
\includegraphics[width = \columnwidth]{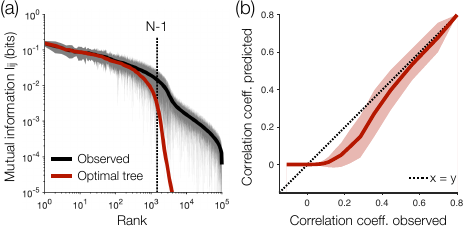} \\
\caption{Predicting pairwise statistics. (a) Ranked order of significant mutual information in the population (black), two--standard--deviation errors (shaded region), and prediction of the minimax entropy model (red). (b) Correlation coefficients predicted in the model versus those in the data, with dashed lines indicating equality. All pairs are divided evenly into bins along the x--axis, with solid lines and shaded regions reflecting means and errors (standard deviations) within bins. \label{fig_corr}}
\end{figure}

While each model $P_\mathcal{T}$ is defined to match a sparse subset of the observed correlations $\left<x_ix_j\right>$ (and thus mutual information $I_{ij}$) for $(ij)\in\mathcal{T}$, we can ask what $P_\mathcal{T}$ predicts for all pairs of neurons (see Appendix \ref{app_corr}). We note that the optimal tree does not simply match the largest $N-1$ values of $I_{ij}$; in general, these will form loops. Yet we find that the minimax entropy model still predicts the distribution of $I_{ij}$ within experimental error for the top $\sim N$ values [Fig.~\ref{fig_corr}(a)]. Indeed, we find that the model captures the strong correlations in the population [Fig.~\ref{fig_corr}(b)]; this accuracy decreases significantly for the minimum distance and random trees (see Appendix \ref{app_comp}). As expected, the optimal tree underpredicts the strengths of weak and negative correlations [Fig.~\ref{fig_corr}(b)]. Although these correlations may seem unimportant individually, we note that they comprise the vast majority of neuron pairs [Fig.~\ref{fig_data}(c)].

\begin{figure}[t]
\centering
\includegraphics[width = .9\columnwidth]{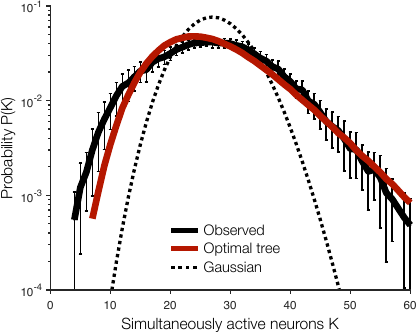} \\
\caption{Predicting synchronized activity. Distribution $P(K)$ of the number of simultaneously active neurons $K$ in the data (black), predicted by the minimax entropy tree (red), and the Gaussian distribution for independent neurons with mean and variance $\langle K\rangle_\text{exp}$ (dashed). We confirm that the Gaussian matches the data after shuffling the activity of each neuron in time. To estimate $P(K)$ and error bars (two standard deviations), we first split the data into 1--minute blocks to preserve dependencies between consecutive samples. We then select one third of these blocks at random and repeat 100 times. For each subsample, we compute the optimal tree $\mathcal{T}$ and predict $P(K)$ using a Monte Carlo simulation of the model $P_\mathcal{T}$. \label{fig_synchrony}}
\end{figure}

With knowledge of only $2/N\sim 0.1\%$ of the pairwise correlations, can the optimal tree capture collective behavior in the system? In neuronal populations (and other complex systems), one key collective property is synchronized activity \cite{Schneidman-01, Tkacik-03, Meshulam-02, Lynn-04}, which is characterized by the probability $P(K)$ that $K$ out of the $N$ neurons are simultaneous active. If the neurons were independent, this distribution would be approximately Gaussian at large $N$ (Fig.~\ref{fig_synchrony}, dashed). But in real populations, the dependencies among neurons leads to a much broader distribution (Fig.~\ref{fig_synchrony}, black), with moments of extreme synchrony in both activity (large $K$) and silence (small $K$). If one builds a model from pairwise correlations chosen at random, then the distribution $P(K)$ is almost indistinguishable from that of an independent system (see Appendix \ref{app_comp}). By contrast, the optimal tree captures most of this collective behavior \cite{Lynn-15}, correctly predicting $\gtrsim 100$--fold increases in the probabilities that $K\gtrsim 50$ neurons are active in the same small time bin (Fig.~\ref{fig_synchrony}, red). Although the detailed patterns of activity in the system are shaped by competing interactions that are missing from our optimal tree, this shows that large--scale synchrony can emerge from a sparse network of the strongest correlations.

\section{Structure of minimax entropy model}
\label{sec_model}

\subsection{Ising structure}

\begin{figure}[b]
\centering
\includegraphics[width = \columnwidth]{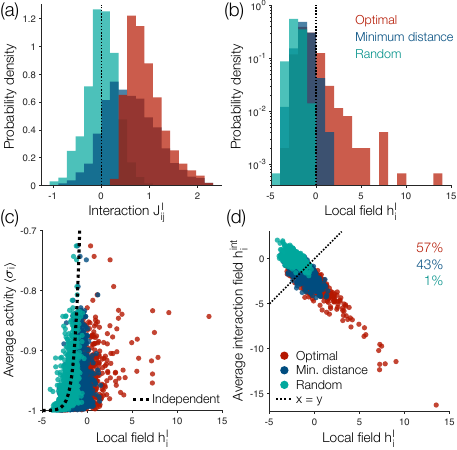} \\
\caption{Maximum entropy models of large--scale activity. (a,b) Distributions of Ising interactions $J^\text{I}_{ij}$ (a) and local fields $h^\text{I}_i$ (b) in the optimal tree (red), the minimum distance tree (blue), and a random tree (cyan). (c) Average activities $\left<\sigma_i\right>$ versus local fields $h^\text{I}_i$, where each point represents an individual neuron. Dashed line illustrates the independent prediction $\left<\sigma_i\right> = \tanh h^\text{I}_i$. (d) Average interaction fields $h^\text{int}_i = \sum_j J^\text{I}_{ij}\left<\sigma_j\right>$ versus local fields $h^\text{I}_i$. Percentages indicate the proportion of neurons for which $h_i^\text{int} < h_i^\text{I}$ (dashed line indicates equality). \label{fig_model}}
\end{figure}

\begin{figure*}[t]
\centering
\includegraphics[width = .8\textwidth]{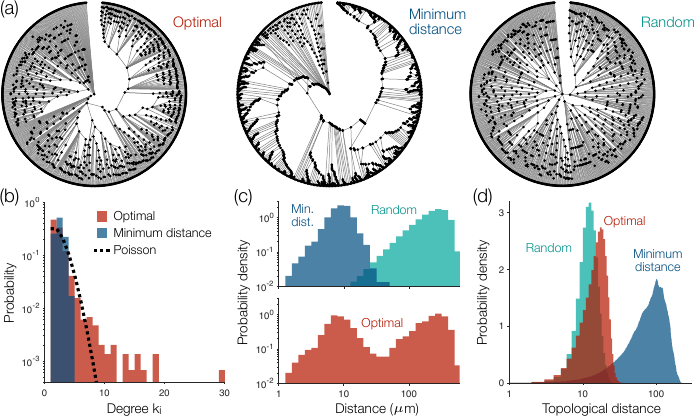} \\
\caption{Network structure of optimal tree. (a) Illustrations of the minimax entropy tree (left), the minimum distance tree (middle), and a random tree (right). In all networks, the central neuron has the largest degree (number of connections), and those with one connection are located on the perimeter with distance from the central neuron decreasing in the clockwise direction. (b) Distributions of degrees in the optimal and minimum distance trees. Dashed line indicates the Poisson distribution of random trees. (c) Distributions of physical distances among the connections in each tree. (d) Distributions of topological distances (or the lengths of shortest paths) across all pairs of neurons in each tree. \label{fig_structure}}
\end{figure*}

To understand the nature of the optimal tree, we can study the minimax entropy model $P_\mathcal{T}$ itself, which, as discussed above, is equivalent to an Ising model from statistical physics. This mapping is made concrete by considering a system of spins $\sigma_i = 2x_i - 1 \in \pm 1$ with Ising interactions $J^\text{I}_{ij} = J_{ij}/4$ and local fields $h^\text{I}_i = h_i/2 + \sum_j J^\text{I}_{ij}$, where $J_{ij}$ and $h_i$ are defined in Eqs. (\ref{eq_J}) and (\ref{eq_h}). If the interaction $J^\text{I}_{ij}$ is positive (negative), then activity in neuron $i$ leads to activity (silence) in neuron $j$, and vice versa. For random trees, the interactions $J^\text{I}_{ij}$ are nearly evenly split between positive and negative [Fig.~\ref{fig_model}(a)]; this is consistent with previous investigations of fully--connected models in populations of $N\sim 100$ neurons \cite{Schneidman-01, Tkacik-03, Meshulam-02}. Meanwhile, we recall that the largest mutual information in the population belongs to positively correlated neurons [Fig.~\ref{fig_data}(d)]. Accordingly, the optimal tree has interactions that are almost exclusively positive [Fig.~\ref{fig_model}(a)]. We have arrived, perhaps surprisingly, at a traditional Ising ferromagnet.

While the interactions $J^\text{I}_{ij}$ define effective influences between neurons, the local fields represent individual biases toward activity ($h^\text{I}_i > 0$) or silence ($h^\text{I}_i < 0$). For random trees, all of the local fields are negative [Fig.~\ref{fig_model}(b)], reflecting the fact that neurons are more likely to be silent than active. But in the optimal tree, we see that some neurons are counterintuitively biased toward activity with $h^\text{I}_i > 0$ [Fig.~\ref{fig_model}(b)]. These positive biases stand in competition with the positive interactions in the model, which, because neurons favor silence, tend to induce silence in the population.

To understand the effects of interactions on individual cells, we note that the average activity of an independent neuron $i$ is fully defined by $h^\text{I}_i$ through the relation $\left<\sigma_i\right> = \tanh h^\text{I}_i$. Since random trees contain only weak correlations, the neuronal activity closely tracks this independent prediction [Fig.~\ref{fig_model}(c)]. As interactions become increasingly positive in the minimum distance and optimal trees, the alignment of neighboring neurons produces average activities that are significantly lower than one would expect from local fields alone [Fig.~\ref{fig_model}(c)]. For each neuron $i$, the competition between internal biases and interactions is made clear by comparing the local field $h^\text{I}_i$ to the average influence due to interactions $h^\text{int}_i = \sum_j J^\text{I}_{ij}\left<\sigma_j\right>$. In random trees, only $1\%$ of neurons are dominated by interactions, such that $h^\text{int}_i < h^\text{I}_i$ [Fig.~\ref{fig_model}(d)]; this proportion increases to $43\%$ in the minimum distance tree and $57\%$ in the optimal tree [Fig.~\ref{fig_model}(d)]. So despite the fact that the tree structure constrains each cell to only interact with two others in the entire population (on average), most neurons in the optimal tree are driven more strongly by interactions than internal biases.

\subsection{Network structure}

In addition to the functional properties of the model $P_\mathcal{T}$, we can also study the graph structure of the optimal tree $\mathcal{T}$. To visualize each tree, we place the cell with the most connections (or largest degree $k_i=|\mathcal{N}_i|$) at the center and all of the cells with single connections ($k_i = 1$) around the perimeter [Fig.~\ref{fig_structure}(a)]. For random trees, the distribution of degrees is Poisson [Fig.~\ref{fig_structure}(b)], preventing the emergence of high--degree hub nodes. Degrees are even more sharply peaked in the minimum distance tree, such that we do not observe a single neuron with more than four connections [Fig.~\ref{fig_structure}(b)]. By contrast, the optimal tree has a much broader degree distribution, with a central neuron that connects to 29 other cells in the population [Fig.~\ref{fig_structure}(b)]. Such hub nodes are frequently observed in the brain's physical connectivity \cite{Song-01, Lynn-05, Lin-02}, and are thought to play an important role in facilitating communication \cite{Albert-01}.

By maximizing information about the population, one might hope that the optimal tree captures features of the true interactions between neurons. In the brain, demands on communication are constrained by energetic costs \cite{Harris-01}. Networks have evolved to balance efficient communication (minimizing the number of steps between cells in the network, known as topological distance) with energetic efficiency (minimizing the physical lengths of connections) \cite{Laughlin-01, Lynn-05}. These pressures are in direct competition: Networks with physically local connections form lattice--like structures with long topological distances, and networks with short topological distances (known as the small--world property \cite{Watts-01}) require physically long--range connections. Indeed, in the minimum distance tree, which is composed of the physically shortest connections [Fig.~\ref{fig_structure}(c)], communication between two neurons requires $\sim100$ intermediate cells on average [Fig.~\ref{fig_structure}(d)]; and random trees, which are known to produce short topological distances [Fig.~\ref{fig_structure}(d)], are mostly composed of long--range connections [Fig.~\ref{fig_structure}(c)]. Meanwhile, the minimax entropy model identifies connections that are much shorter than average [Fig.~\ref{fig_structure}(c)] while simultaneously maintaining small--world structure [Fig.~\ref{fig_structure}(d)], just as observed in real neuronal networks \cite{Laughlin-01}.

\section{Scaling with population size}
\label{sec_scaling}

\begin{figure}[t]
\centering
\includegraphics[width = \columnwidth]{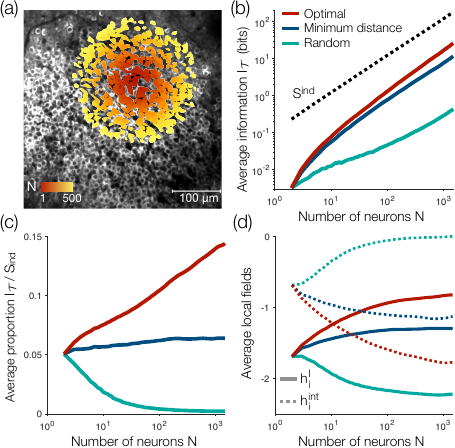} \\
\caption{Scaling of the minimax entropy model. (a) Illustration of our growth process superimposed on a fluorescence image of the $N = 1485$ neurons in the mouse hippocampus. Starting with a single neuron $i$, we grow the population of $N$ neurons closest to $i$ (red to yellow). We then repeat this process for each neuron and average the results. (b) Information $I_\mathcal{T}$ captured by different trees as a function of population size $N$. Dashed line indicates the independent entropy $S_\text{ind}$. (c) Fraction of the independent entropy $I_\mathcal{T}/S_\text{ind}$ explained by different trees as a function of $N$. (d) Average Ising local fields $h^\text{I}_i$ (solid) and interaction fields $h^\text{int}_i$ (dashed) for the maximum entropy models on different trees $P_\mathcal{T}$. \label{fig_scaling}}
\end{figure}

Thus far, we have focused on a single population of $N \sim 1500$ neurons. But as experiments advance to record from even larger populations, how does the minimax entropy model scale with $N$? To answer this question, in the spirit of Ref.~\cite{Meshulam-02} one can imagine growing a contiguous population centered at a single neuron [Fig.~\ref{fig_scaling}(a)], and computing the optimal tree for increasing population sizes. Due to the efficiency of our model, we can repeat this process starting from each of the different neurons and average over the results. As the population grows, the independent entropy $S_\text{ind}$ must increase extensively (that is, linearly with $N$) on average [Fig.~\ref{fig_scaling}(b)]. Since each tree contains $N-1$ correlations, one might also expect the information $I_\mathcal{T}$ of any tree to scale extensively. However, we find that the scaling of $I_\mathcal{T}$ with population size depends critically on which correlations we use in building the tree
%the nature of the correlations 
[Fig.~\ref{fig_scaling}(b)].

If the information $I_\mathcal{T}$ grows extensively, then the model $P_\mathcal{T}$ explains a constant proportion of the independent entropy $I_\mathcal{T}/S_\text{ind}$ across different population sizes. Indeed, because the properties of the closest neurons do not change as the population grows (on average), the minimum distance tree captures a nearly constant $\sim6\%$ of the independent entropy [Fig.~\ref{fig_scaling}(c)]. By contrast, since the mutual information $I_{ij}$ between neurons tends to decrease with physical distance [Fig.~\ref{fig_data}(f)], the average mutual information $\bar{I}$ in a spatially contiguous population decreases with $N$. Thus, the typical information in a random tree %$I_\mathcal{T} \approx (N-1)\bar{I}$ 
grows subextensively with the population size [Fig.~\ref{fig_scaling}(b)], and the fractional information $I_\mathcal{T}/S_\text{ind}$ vanishes [Fig.~\ref{fig_scaling}(c)].

But even though the average mutual information $\bar{I}$ decreases, as the population grows we uncover more of the exceptionally large mutual information $I_{ij}$ in the tail of the distribution [Fig.~\ref{fig_data}(b)]. By identifying these highly informative correlations, the optimal tree accumulates a superextensive amount of information $I_\mathcal{T}$ [Fig.~\ref{fig_scaling}(b)], thus capturing a greater proportion of the independent entropy as $N$ increases [Fig.~\ref{fig_scaling}(c)]. This increased explanatory power is underpinned by stronger interactions and weaker local fields in the Ising network [Fig.~\ref{fig_scaling}(d)]. There is no sign that the trend in Fig.~\ref{fig_scaling}(c) is saturating at $N\sim 10^3$, suggesting that our minimax entropy framework may become even more effective for larger populations.

\section{Thermodynamics and signatures of criticality}
\label{sec_thermo}

\begin{figure*}[t]
\centering
\includegraphics[width = .85\textwidth]{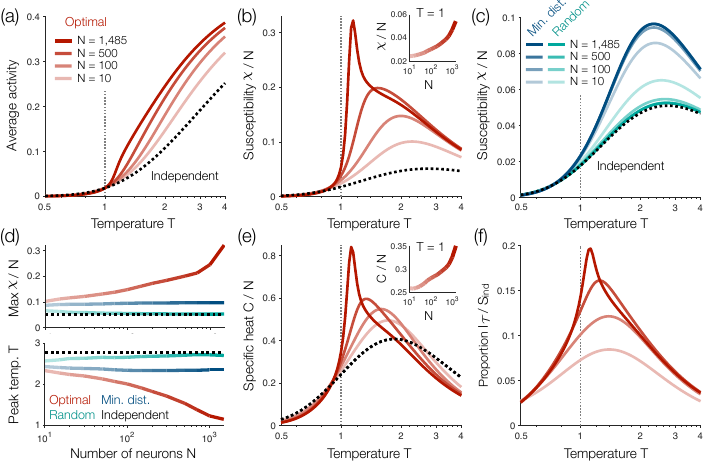} \\
\caption{Thermodynamics of the minimax entropy model. (a) Average activity $\frac{1}{N}\sum_i \left<x_i\right>_T$ as a function of temperature $T$ in the optimal tree $\mathcal{T}$ for populations of increasing size $N$ and for independent neurons (dashed). (b) Normalized susceptibility $\chi/N$ versus temperature $T$ for the same models as (a). Inset shows the increase in $\chi/N$ with population size $N$ for the true minimax entropy models ($T = 1$). (c) Normalized susceptibility versus $T$ for the minimum distance tree (blue) and random trees (cyan) across different population sizes. (d) Maximum value of the normalized susceptibility $\chi/N$ (top) and peak temperature $T$ (bottom) as functions of the population size $N$ for the optimal tree (red), the minimum distance tree (blue), random trees (cyan), and independent neurons (dashed). (e,f) Specific heat $C/N$ (e) and information fraction $I_\mathcal{T}/S_\text{ind}$ (f) versus temperature $T$ for the minimax entropy models in (a) and (b). Inset in (e) shows the increase in $C/N$ with population size $N$ for the true minimax entropy models ($T = 1$). In all panels, darker lines reflect populations of increasing size $N$, constructed using the method in Fig.~\ref{fig_scaling}(a) and averaged over 100 random initializations. In (a)--(e), dashed lines represent independent neurons. \label{fig_thermo}}
\end{figure*}

As discussed above, each tree of observed correlations $\mathcal{T}$ generates a maximum entropy model $P_\mathcal{T}$ [Eq.~(\ref{eq_PG})], which in turn is equivalent to a system of Ising spins. This mapping from experimental observations to statistical physics gives us the opportunity to ask whether the model $P_\mathcal{T}$ occupies a special place in the space of possible models. In statistical mechanics, equilibrium systems are described by the Boltzmann distribution,
\begin{equation}
\label{eq_Bolt}
P(\bm{x}) = \frac{1}{Z}\exp\left[-\frac{1}{T}E(\bm{x})\right],
\end{equation}
where $T$ is the temperature of the system and $E(\bm{x})$ is the Hamiltonian, which defines the energy of state $\bm{x}$. For a given tree $\mathcal{T}$, we notice that $P_\mathcal{T}$ defines a Boltzmann distribution with temperature $T = 1$ and energy
\begin{equation}
\label{eq_E}
E(\bm{x}) = -\sum_{(ij)\in \mathcal{T}} J_{ij}x_ix_j - \sum_i h_ix_i,
\end{equation}
where $J_{ij}$ and $h_i$ are defined in Eqs. (\ref{eq_J}) and (\ref{eq_h}). Note that we do not assume the experimental system itself is in equilibrium; this correspondence is purely mathematical.

By perturbing the temperature away from $T = 1$, we can probe at least one slice through the space of possible networks \cite{Tkacik-03}. For each value of $T$, we arrive at a hypothetical system $P_\mathcal{T}(\bm{x};T)$ with average activities $\left<x_i\right>_T$ and correlations $\left<x_ix_j\right>_T$ that are no longer constrained to match experimental observations. Consider the minimax entropy tree, which (as discussed in \S\ref{sec_model}) produces a ferromagnetic Ising model with nearly all positive interactions $J^\text{I}_{ij}$ [Fig.~\ref{fig_model}(a)]. At high temperatures $T \gg 1$, fluctuations destroy the preference for silence over activity, and the system approaches the average activity $\frac{1}{N}\sum_i \left<x_i\right>_T = 0.5$ [Fig.~\ref{fig_thermo}(a)]. Meanwhile, at low temperatures $T \ll 1$, activity vanishes as the network freezes into the all--silent ground state $\bm{x} = 0$ [Fig.~\ref{fig_thermo}(a)]. In both limits, all of the information contained in correlations is lost.

As the temperature decreases, most systems experience a gradual transition from disorder to order. But for certain combinations of parameters $J_{ij}$ and $h_i$, a small change in the temperature $T$ can lead to a large change in the behavior of the system, and as $N$ becomes large, this transition becomes sharp \cite{sethna_06, Tkacik-03, Schnabel-01}. Such phase transitions mark a critical point in the space of possible systems, with the Ising ferromagnet as the canonical example \cite{Peierls-01}. In the optimal tree, as the temperature increases just above $T = 1$, the positive interactions lead to a much steeper increase in activity than an independent system; and this transition grows even sharper for larger populations [Fig.~\ref{fig_thermo}(a)].

We emphasize that at any finite $N$ there is no true critical point, but $N\sim 1000$ may be large enough that the idealization $N\rightarrow\infty$ is useful.  Since we are studying models defined on trees, there are also subtleties about how one would construct the thermodynamic limit, since such a large fraction of sites are on the boundary \cite{Baxter-01}.  For our purposes, the interesting question is whether real networks of neurons are in any sense at special points in the space of possible networks.  One way in which this could happen is if parameters are set so that simple macroscopic quantities have near--extremal values.

One example of a macroscopic quantity that provides a global measure of collective behavior is the total susceptibility of the mean activity to changes in the bias fields,
%In statistical physics, critical points are often marked by a divergence in the susceptibility of the order parameter. Here, the order parameter is the average activity, so the susceptibility is the total connected correlation
\begin{equation}
\label{eq_chi}
\chi = \sum_{ij} \frac{d \left<x_i\right>_T}{d h_j} = \frac{1}{T}\sum_{ij}\left(\left<x_ix_j\right>_T - \left<x_i\right>_T\left<x_j\right>_T\right),
\end{equation}
where the rewriting in terms of connected correlations can be derived from the Boltzmann distribution [Eq.~(\ref{eq_Bolt})]. We recall that at conventional critical points we would see a divergence of $\chi/N$ as $N\rightarrow\infty$.  At both high and low temperatures, correlations are destroyed, and the susceptibility vanishes [Fig.~\ref{fig_thermo}(b)]. However, at intermediate temperatures, the susceptibility exhibits a peak that becomes sharper as the system grows, even after normalizing by the population size $N$ [Fig.~\ref{fig_thermo}(b)]. Moreover, as $N$ increases, the peak temperature decreases toward $T = 1$, corresponding to the true minimax entropy model $P_\mathcal{T}$. By contrast, the minimum distance and random trees undergo smooth transitions from disorder to order [Fig.~\ref{fig_thermo}(c)], with the maximum susceptibility and peak temperatures remaining approximately constant across all population sizes $N$ [Fig.~\ref{fig_thermo}(d)].

In addition to the susceptibility $\chi$, we also observe a dramatic peak in the specific heat $C/N$ [Fig.~\ref{fig_thermo}(e)], where
\begin{equation}
\label{eq_C}
C = \frac{d \left<E(\bm{x})\right>_T}{d T}
\end{equation}
is the heat capacity (see Appendix \ref{app_thermo}). Although there is no meaning to ``heat'' in this system, because the specific heat is related to the variance in energy, we can think of the peak in $C$ as being a peak in the dynamic range of (log) probabilities across the states of the network. These divergences in the susceptibility and heat capacity also align with a sharp peak in the information fraction $I_\mathcal{T}/S_\text{ind}$ [Fig.~\ref{fig_thermo}(f)], with larger systems becoming even more strongly correlated. Together, these results indicate that the true minimax entropy model $P_\mathcal{T}$ is poised near a special point in the space of models $P_\mathcal{T}(\bm{x};T)$, where small changes in parameter values can produce large changes in the collective behavior of the system.

\section{Conclusions}
\label{sec_conclusions}

The maximum entropy principle provides the most unbiased mapping from experimental observations to statistical physics models. Over the past two decades, this link has proven useful in understanding the emergence of collective behaviors in populations of neurons and other complex living systems \cite{Schneidman-01, Nguyen-01, Meshulam-17, Meshulam-02, Tkacik-03, Lezon-01, weigt+al_09, marks2011protein, lapedes2012using, Bialek-01, russ2020evolution, Lynn-04}. Less widely emphasized is the fact that there is not a single maximum entropy model, but rather a landscape of possible models depending on what features of the system we choose to constrain. Quite generally, we should choose the features that are most informative---the ones that minimize the entropy of the maximum entropy model---leading to the minimax entropy principle \cite{Zhu-01}. As experiments record from larger and larger populations of neurons \cite{segev+al2004, litke+al2004, chung2019high, dombeck2010functional, tian2012neural, demas+al2021, steinmetz2021neuropixels}, we enter an undersampled regime in which selecting a limited number of maximally informative features is not only conceptually appealing, but also practically necessary.

While the minimax entropy problem is generally intractable, here we make progress in two steps. First, we build upon previous work by constraining mean activities and pairwise correlations, resulting in models that are equivalent to systems of Ising spins. Second, taking inspiration from the Bethe lattice, we focus only on trees of correlations, or sparse networks without loops. Under these restrictions, we solve the minimax entropy problem exactly, identifying the optimal tree in quadratic time \cite{Chow-01, Nguyen-01}. The result is a non--trivial family of statistical physics models that can be constructed very efficiently for large neuronal populations.

It is far from obvious that these models can capture any of the essential collective behavior in real networks. To answer this question, we study a population of $N\sim 1500$ neurons in the mouse hippocampus \cite{Gauthier-01}, identifying the maximally informative tree of pairwise correlations (Figs.~\ref{fig_model} and \ref{fig_structure}). Despite containing only one correlation per neuron, this minimax entropy model accounts for $14\%$ of the independent entropy (over $50$ times more than random trees) and predicts the distribution of large--scale synchrony in activity (Fig.~\ref{fig_synchrony}). Moreover, the model becomes more effective as the population grows (Fig.~\ref{fig_scaling}) and exhibits hints of critical behavior (Fig.~\ref{fig_thermo}). The success of such a sparse model hinges on the fact that the distribution of mutual information between neurons is heavy--tailed [Fig.~\ref{fig_data}(b)], such that a few rare correlations carry much more information than average. In fact, the physical connections between neurons are now understood to be heavy--tailed across a range of animals \cite{Lynn-13}, suggesting that our approach my prove effective in other neural systems. While these minimax entropy models cannot capture all of a system's collective properties, they provide at least a starting point for simplified descriptions of the much larger systems becoming accessible in modern experiments.

\begin{acknowledgments}
We thank L.\ Meshulam and J.L.\ Gauthier for guiding us through the data of Ref.~\cite{Gauthier-01}, and C.M.\ Holmes and D.J.\ Schwab for helpful discussions. This work was supported in part by the National Science Foundation through the Center for the Physics of Biological Function (PHY--1734030), by the National Institutes of Health through the BRAIN initiative (R01EB026943), by the James S McDonnell Foundation through a Postdoctoral Fellowship Award (C.W.L.), and by Fellowships from the Simons Foundation and the John Simon Guggenheim Memorial Foundation (W.B.).
\end{acknowledgments}

\appendix

\section{Ising calculations on a tree}
\label{app_Ising}

To establish notation, we begin by reviewing well known ideas about statistical mechanics for models without loops.  We then proceed, here and in subsequent Appendices, to technical points needed for the main text.

\subsection{Partition function}

Consider a system of $N$ binary variables $x_i \in \{0,1\}$, $i = 1,2,\hdots,N$, defined by fields $h_i$ and interactions $J_{ij}$ that lie on a tree $\mathcal{T}$. The Boltzmann distribution [Eq.~(\ref{eq_PG})] takes the form
\begin{equation}
\label{eq_PT}
P_\mathcal{T}(\bm{x}) = \frac{1}{Z}\text{exp}\Bigg[\sum_{(ij)\in \mathcal{T}} J_{ij}x_ix_j + \sum_i h_ix_i - F\Bigg],
\end{equation}
where $F = 0$ is the zero--point energy, which will become useful. To begin, we seek to compute the partition function,
\begin{equation}
\label{eq_Z}
Z = \sum_{\bm{x}} \text{exp}\Bigg[\sum_{(ij)\in \mathcal{T}} J_{ij}x_ix_j + \sum_i h_ix_i - F\Bigg].
\end{equation}
To do so, imagine summing over one variable, and finding a new system of $N-1$ variables with the same partition function $Z$. If we can repeat this process until no variables remain, then computing $Z$ will be trivial.

We label the nodes $i$ based on the order that they are removed, and we let $h^{(i)}_i$ and $F^{(i)}$ denote the updated parameters at step $i$, while the interactions $J_{ij}$ stay fixed. Consider summing over a variable $i$ with only one connection in the network, say to variable $j$. We note that such a node is always guaranteed to exist in a tree. To keep the partition function fixed, the new system with $i$ removed must satisfy the equations
\begin{equation}
\label{eq_sum}
e^{h_j^{(i)}x_j - F^{(i)}}\big(e^{J_{ij}x_j + h^{(i)}_i }+ 1\big) = e^{h_j^{(i+1)}x_j - F^{(i+1)}}.
\end{equation}
This is a system of two equations (one for each value of $x_j$), which we can solve for the new parameters
\begin{align}
\label{eq_hj}
h_j^{(i+1)} &= h_j^{(i)} + \ln\Bigg[\frac{e^{J_{ij} + h_i^{(i)}} + 1}{e^{h_i^{(i)}} + 1} \Bigg], \\
\label{eq_F}
F^{(i+1)} &= F^{(i)} - \ln\big[e^{h_i^{(i)}} + 1 \big].
\end{align}
After removing $i$, the new system still forms a tree, so we can repeat the above procedure. When all nodes have been removed, we are left with a single parameter $\mathcal{F} = F^{(N+1)}$, which is the free energy of the system, and the partition function is given by
\begin{equation}
Z = e^{-\mathcal{F}}.
\end{equation}

\subsection{Average activities and correlations}

To compute population statistics, one simply needs to take derivatives of the partition function,
\begin{align}
\langle x_i\rangle &= \frac{d \ln Z}{d h_i} = -\frac{d\mathcal{F}}{d h_i}, \\
\langle x_ix_j\rangle &= \frac{d \ln Z}{d J_{ij}} = -\frac{d\mathcal{F}}{d J_{ij}},
\end{align}
where $\frac{d}{d h_i}$ and $\frac{d}{d J_{ij}}$ represent total derivatives, which account for indirect dependencies via Eqs. (\ref{eq_hj}) and (\ref{eq_F}). Since $\frac{d\mathcal{F}}{d F^{(i+1)}} = 1$ and $\frac{d h_i^{(i)}}{d h_i} = 1$, the above procedure yields
\begin{equation}
\langle x_i\rangle = -\frac{\partial F^{(i+1)}}{\partial h^{(i)}_i} - \frac{d \mathcal{F}}{d h^{(i+1)}_j}\frac{\partial h^{(i+1)}_j}{\partial h^{(i)}_i}.
\end{equation}
Noticing that
\begin{equation}
- \frac{d \mathcal{F}}{d h^{(i+1)}_j} = - \frac{d \mathcal{F}}{d h_j} = \langle x_j\rangle,
\end{equation}
and taking derivatives of Eqs. (\ref{eq_hj}) and (\ref{eq_F}), we have
\begin{align}
\label{eq_xi}
\langle x_i\rangle &= \frac{1}{1 + e^{-h_i^{(i)}}} \\
&\quad\quad + \langle x_j\rangle \Bigg( \frac{1}{1 + e^{-J_{ij} - h_i^{(i)}}} - \frac{1}{1 + e^{-h_i^{(i)}}}\Bigg). \nonumber
\end{align}
The correlation follows analogously,
\begin{equation}
\label{eq_xixj}
\langle x_ix_j\rangle = - \frac{d \mathcal{F}}{d h^{(i+1)}_j}\frac{\partial h^{(i+1)}_j}{\partial J_{ij}} = \frac{\langle x_j\rangle}{1 + e^{-J_{ij} - h_i^{(i)}}}.
\end{equation}
Thus, by proceeding in the opposite order from which the nodes were removed, we can compute the average activities $\langle x_i\rangle$ and correlations $\langle x_ix_j\rangle$ for $(ij)\in \mathcal{T}$. For the correlations $\langle x_ix_j\rangle$ off the tree (that is, for $(ij) \not\in \mathcal{T}$), see Appendix \ref{app_corr}.

\section{Maximum entropy on a tree}
\label{app_maxent}

We now solve the inverse problem for the parameters $h_i$ and $J_{ij}$ given the observations $\langle x_i\rangle$ and $\langle x_ix_j\rangle$ on a tree $\mathcal{T}$. Inverting Eqs. (\ref{eq_xi}) and (\ref{eq_xixj}) yields
\begin{align}
h_i^{(i)} &= \ln \left[ \frac{\langle x_i\rangle - \langle x_ix_j\rangle}{1 - \langle x_i\rangle - \langle x_j\rangle + \langle x_ix_j\rangle}\right] \\
J_{ij} &= \ln \left[\frac{\langle x_ix_j\rangle}{\langle x_j\rangle - \langle x_ix_j\rangle}\right] - h_i^{(i)}.
\end{align}
Combining the above equations, we can solve for the interaction $J_{ij}$ in Eq.~(\ref{eq_J}). To compute the local field $h_i$, we note that we can repeat the procedure in Appendix \ref{app_Ising} ending at any node; this is equivalent to choosing the root of the tree. If we choose $i$ to be the final node, then we have
\begin{equation}
h_i^{(N)} = \ln \frac{\langle x_i\rangle}{1 - \langle x_i\rangle}.
\end{equation}
Additionally, for each neighbor $j \in \mathcal{N}_i$, Eq.~(\ref{eq_hj}) tells us that we receive a contribution to $h_i^{(N)}$ of the form
\begin{align}
h_i^{(j+1)} - h_i^{(j)} &= \ln\Bigg[\frac{e^{J_{ij} + h_j^{(j)}} + 1}{e^{h_j^{(j)}} + 1} \Bigg] \\
&= \ln \left[\frac{\langle x_i\rangle\left(1 - \langle x_i\rangle - \langle x_j\rangle + \langle x_ix_j\rangle\right)}{\left(1 - \langle x_i\rangle\right)\left(\langle x_i\rangle - \langle x_ix_j\rangle\right)}\right]. \nonumber
\end{align}
Combining these contributions yields
\begin{align}
h_i &= h_i^{(N)} - \sum_{j\in\mathcal{N}_i} \big(h_i^{(j+1)} - h_i^{(j)}\big) \\
&= \ln \frac{\langle x_i\rangle}{1 - \langle x_i\rangle} \\
& \quad\quad + \sum_{j \in \mathcal{N}_i} \ln \left[\frac{\left(1 - \langle x_i\rangle\right)\left(\langle x_i\rangle - \langle x_ix_j\rangle\right)}{\langle x_i\rangle\left(1 - \langle x_i\rangle - \langle x_j\rangle + \langle x_ix_j\rangle\right)}\right]. \nonumber
\end{align}
We have thus arrived at an analytic solution to the maximum entropy problem on a tree.

\section{Information in a tree of correlations}
\label{app_info}

Our ability to efficiently construct the optimal tree $\mathcal{T}$ depends critically on the decomposition of the information $I_\mathcal{T}$ into the sum of mutual information $I_{ij}$ over pairs $(ij)\in \mathcal{T}$ [Eq.~(\ref{eq_I})]. To derive this result, we note that for each connection $(ij)\in \mathcal{T}$, the observables $\langle x_i\rangle$, $\langle x_j\rangle$, and $\langle x_ix_j\rangle$ fully define the marginal distribution $P_{ij}(x_i,x_j)$. Now consider a new tree $\mathcal{T'} = \mathcal{T}/(ij)$ with the connection $(ij)$ removed, such that we do not observe $\langle x_ix_j\rangle$. Since $\mathcal{T}$ has no loops, after removing $(ij)$ the two elements $i$ and $j$ become independent. Meanwhile, the dependence of the rest of the system on $i$ and $j$ remains fixed. Thus, observing the correlation $\langle x_ix_j\rangle$ leads to a drop in entropy
\begin{equation}
\label{eq_Iij}
S_\mathcal{T'} - S_\mathcal{T} = S(P_i) + S(P_j) - S(P_{ij}) = I_{ij},
\end{equation}
where $I_{ij}$ is the observed mutual information between $i$ and $j$. Repeating the above argument for every correlation in $\mathcal{T}$, we arrive at Eq.~(\ref{eq_I}).

\section{Estimating mutual information}
\label{app_MI}

In order to estimate the mutual information between neurons $I_{ij}$, one must correct for finite--data effects \cite{Strong-01}. To do so, we subsample the data hierarchically for different data fractions $\{1,0.9,\hdots,0.2,0.1\}$, such that each subsample is contained within the larger subsamples. Additionally, to preserve the dependencies between consecutive data points $\bm{x}^{(m)}$ and $\bm{x}^{(m+1)}$, we sample temporally contiguous fractions of the data. To ensure that each point is sampled with equal probability, we allow subsamples that span the beginning and end of the recording.

\begin{figure}[t]
\centering
\includegraphics[width = .9\columnwidth]{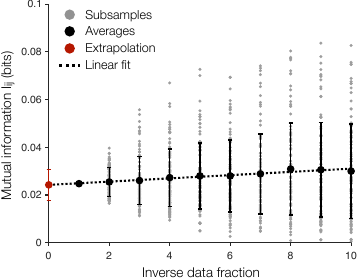} \\
\caption{Correcting for finite--data effects on mutual information. For a given pair of neurons $i$ and $j$, we plot the estimated mutual information $I_{ij}$ [Eq.~(\ref{eq_Iest})] versus the inverse data fraction for individual subsamples of the data (grey). Repeating 100 times, we plot the average $I_{ij}$ for each data fraction (black), a linear fit (dashed line), and the infinite--data estimate (red). Data points and error bars reflect means and standard deviations over the 100 repetitions. \label{fig_finite}}
\end{figure}

For each subsample, we estimate the mutual information between neurons $i$ and $j$ using the equation
\begin{equation}
\label{eq_Iest}
I_{ij} = \sum_{x_i,x_j} \tilde{P}_{ij}(x_i,x_j) \log \frac{\tilde{P}_{ij}(x_i,x_j)}{\tilde{P}_i(x_i)\tilde{P}_j(x_j)},
\end{equation}
where
\begin{align}
\label{eq_Pij}
\tilde{P}_{ij}(x_i,x_j) &= \frac{1}{M+1}\Big(1 + \sum_{m = 1}^M \delta_{x_i,x_i^{(m)}}\delta_{x_j,x_j^{(m)}}\Big), \\
\tilde{P}_i(x_i) &= \sum_{x_j} \tilde{P}_{ij}(x_i,x_j).
\end{align}
The pseudo--counts in Eq.~(\ref{eq_Pij}) ensure that the mutual information estimates do not diverge. After estimating $I_{ij}$ for each data fraction, following Ref.~\cite{Strong-01} we extrapolate to the infinite--data limit using a linear fit with respect to the inverse data fraction. Repeating this process 100 times, we arrive at a distribution of infinite--data estimates for $I_{ij}$, from which we can compute a mean and standard deviation (Fig.~\ref{fig_finite}). To check the above procedure, we note that shuffling the activity of each neuron in time should destroy the mutual information $I_{ij}$. Indeed, for time--shuffled data, we do not observe a single significant mutual information in the population.

\section{Computing all correlations}
\label{app_corr}

Given a maximum entropy model with parameters $h_i$ and $J_{ij}$ on a tree $\mathcal{T}$, in Appendix \ref{app_Ising} we showed how to compute the averages $\langle x_i\rangle$ and correlations $\langle x_ix_j\rangle$ on the tree. Specifically, we computed the partition function $Z$ by summing over variables $x_i$ in the order $i = 1,2,\cdots,N$, and then computed statistics in the reverse order. Here, we show how to compute the correlations $\langle x_ix_j\rangle$ not on the tree; that is, for $(ij) \not\in \mathcal{T}$. To begin, we assume that we have computed the correlations $\langle x_j x_k\rangle$ for all nodes $k > i > j$. Then, if we compute $\langle x_ix_j\rangle$, the procedure will follow by induction.

From the Boltzmann distribution in Eq.~(\ref{eq_PG}), we have
\begin{equation}
\frac{d \langle x_i\rangle}{d h_j} = \langle x_ix_j\rangle - \langle x_i\rangle \langle x_j\rangle.
\end{equation}
We already know how to compute the averages $\langle x_i\rangle$ and $\langle x_j\rangle$, so all that remains is to calculate the above derivative. Let $p(i)$ denote the parent of $i$ (that is, the final neighbor when $i$ is removed) and likewise for $p(j)$. Differentiating Eq.~(\ref{eq_xi}) with respect to $h_j$, we have
\begin{equation}
\label{eq_dmdh}
\frac{d \langle x_i\rangle}{d h_j} = \frac{\partial \langle x_i\rangle}{\partial h_i^{(i)}}\frac{d h_i^{(i)}}{d h_j} + \frac{\partial \langle x_i\rangle}{\partial \langle x_{p(i)}\rangle} \frac{d \langle x_{p(i)} \rangle}{d h_j}.
\end{equation}
We note that
\begin{equation}
\frac{d \langle x_{p(i)} \rangle}{d h_j} = \langle x_{p(i)}x_j\rangle - \langle x_{p(i)}\rangle \langle x_j\rangle,
\end{equation}
which we have already computed by assumption, since $p(i) > i$. From Eq.~(\ref{eq_xi}) we have,
\begin{align}
\frac{\partial \langle x_i\rangle}{\partial h_i^{(i)}} &= \frac{e^{-h_i^{(i)}}}{\big(1 + e^{-h_i^{(i)}}\big)^2} \\
&\quad + \langle x_{p(i)}\rangle \Bigg( \frac{e^{-J_{ip(i)} - h_i^{(i)}}}{\big(1 + e^{-J_{ip(i)} - h_i^{(i)}}\big)^2} - \frac{e^{-h_i^{(i)}}}{\big(1 + e^{-h_i^{(i)}}\big)^2}\Bigg), \nonumber
\end{align}
and
\begin{equation}
\frac{\partial \langle x_i\rangle}{\partial \langle x_{p(i)}\rangle} = \frac{1}{1 + e^{-J_{ip(i)} - h_i^{(i)}}} - \frac{1}{1 + e^{-h_i^{(i)}}}.
\end{equation}
Finally, we note that the dependence of $h_i^{(i)}$ on $h_j$ runs only through $h^{(j+1)}_{p(j)}$, such that
\begin{equation}
\frac{d h_i^{(i)}}{d h_j} = \frac{d h_i^{(i)}}{d h^{(j+1)}_{p(j)}} \frac{\partial h^{(j+1)}_{p(j)}}{\partial h_j}.
\end{equation}
Since $p(j) > j$, we can assume that we have already computed $\frac{d h_i^{(i)}}{d h^{(j+1)}_{p(j)}}$. Finally, Eq.~(\ref{eq_hj}) yields
\begin{equation}
\frac{\partial h^{(j+1)}_{p(j)}}{\partial h_j} = \frac{1}{1 + e^{-J_{jp(j)} - h_j^{(j)}}} - \frac{1}{1 + e^{-h_j^{(j)}}}.
\end{equation}
Plugging everything into Eq.~(\ref{eq_dmdh}), and inducting on $i > j$, one can compute the correlations $\langle x_ix_j\rangle$ between all variables.

\section{Predictions of random and minimum distance trees}
\label{app_comp}

\begin{figure}[b]
\centering
\includegraphics[width = \columnwidth]{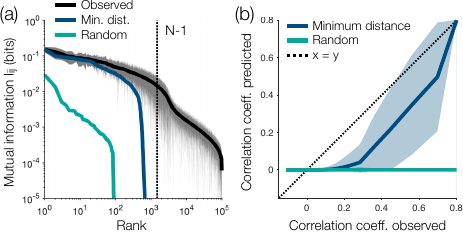} \\
\caption{Pairwise statistics in minimum distance and random trees. (a) Ranked order of significant mutual information in the population (black), two--standard--deviation errors (shaded region), and predictions of the minimum distance tree (blue) and a random tree (cyan). (b) Correlation coefficients predicted in the minimum distance and random tree models versus those in the data, with dashed lines indicating equality. All pairs are divided evenly into bins along the x--axis, with solid lines and shaded regions reflecting means and errors (standard deviations) within bins. \label{fig_corr2}}
\end{figure}

In \S\ref{sec_synchrony}, we studied the predictions of the minimax entropy model $P_\mathcal{T}$ corresponding to the optimal tree $\mathcal{T}$. For comparison, here we consider the predictions of the minimum distance and random trees. While the optimal tree captures $I_\mathcal{T} = 26.2\,\text{bits}$ of information ($I_\mathcal{T}/S_\text{ind} = 14.4\%$ of the independent entropy), the minimum distance tree only captures $11.9\,\text{bits}$ of information ($6.5\%$ of the independent entropy), and a typical random tree only captures $(N-1)\bar{I} = 0.4\,\text{bits}$ ($0.2\%$ of the independent entropy).

\begin{figure}[t]
\centering
\includegraphics[width = .9\columnwidth]{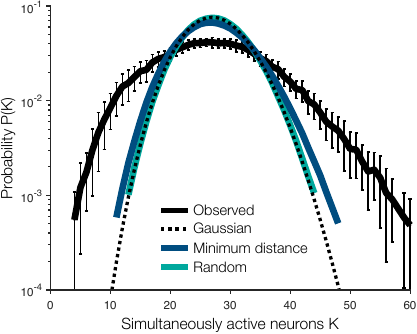} \\
\caption{Synchronized activity in minimum distance and random trees. Distribution $P(K)$ of the number of simultaneously active neurons $K$ in the data (black), the Gaussian distribution for independent neurons with mean and variance $\langle K\rangle_\text{exp}$ (dashed), and the predictions of the minimum distance tree (blue) and a random tree (cyan). As in Fig.~\ref{fig_synchrony}, to estimate $P(K)$ and error bars (two standard deviations), we first split the data into 1--minute blocks to preserve dependencies between consecutive samples. We then select one third of these blocks at random and repeat 100 times. For each subsample of the data, we fit the maximum entropy model $P_\mathcal{T}$ for each tree $\mathcal{T}$ and predict $P(K)$ using a Monte Carlo simulation. \label{fig_synchrony2}}
\end{figure}

For each tree, we can predict the mutual information $I_{ij}$ and correlation coefficients between all pairs of neurons using the procedure in Appendix \ref{app_corr}. Since the minimum distance tree includes some of the largest mutual information in the population, it is able to match the distribution of $I_{ij}$ (within errors) out to $N\sim 100$ neurons [Fig.~\ref{fig_corr2}(a)]. However, the minimum distance tree fails to predict the observed correlations across most of the dynamic range of the data [Fig.~\ref{fig_corr2}(b)]. Meanwhile, random trees typically include only weak mutual information [Fig.~\ref{fig_corr2}(a)], such that their predictions are nearly indistinguishable from a population of independent neurons [Figs.~\ref{fig_corr2}(b)].

For the minimax entropy model, a backbone of strong positive interactions combine to produce accurate predictions for the distribution $P(K)$ of population--wide synchrony $K$ (Fig.~\ref{fig_synchrony}). By contrast, random trees predict a Gaussian distribution consistent with independent neurons, and the minimum distance tree only produces a slightly broader distribution (Fig.~\ref{fig_synchrony2}). In both models, large--scale synchrony in activity ($K\gtrsim 50$) and silence ($K \lesssim 10$) occurs significantly less frequently than observed the data.

\section{Thermodynamic quantities}
\label{app_thermo}

Consider a system with fields $h_i$, interactions $J_{ij}$ that lie on a tree $\mathcal{T}$, and temperature $T$. The Boltzmann distribution $P(\bm{x})$ takes the form in Eq.~(\ref{eq_Bolt}) with energy $E(\bm{x})$ defined in Eq.~(\ref{eq_E}). Here, we denote averages over $P(\bm{x})$ by $\langle \cdot\rangle$, while dropping the subscript $T$. The susceptibility $\chi$ [Eq.~(\ref{eq_chi})], can be computed using the results of Appendix \ref{app_corr}.

To compute the heat capacity $C$ [Eq.~(\ref{eq_C})], we begin with the average energy
\begin{equation}
\label{eq_Eavg}
\langle E(\bm{x})\rangle = T^2\frac{d \ln Z}{d T} = -T^2 \frac{d (\mathcal{F}/T)}{d T} = \mathcal{F} - T \frac{d \mathcal{F}}{d T}.
\end{equation}
To calculate the free energy $\mathcal{F}$, we proceed as in Appendix \ref{app_Ising}. After including the temperature $T$, Eqs. (\ref{eq_hj}) and (\ref{eq_F}) take the form
\begin{align}
\label{eq_hjT}
h_j^{(i+1)} &= h_j^{(i)} +  T\ln\Bigg[\frac{e^{\frac{1}{T}(J_{ij} + h_i^{(i)})} + 1}{e^{\frac{1}{T}h_i^{(i)}} + 1} \Bigg], \\
\label{eq_FT}
F^{(i+1)} &= F^{(i)} - T\ln\big[e^{\frac{1}{T}h_i^{(i)}} + 1 \big].
\end{align}
Iteratively summing over each variable, we arrive at the free energy $\mathcal{F} = F^{(N+1)}$. To compute $\frac{d \mathcal{F}}{d T}$, we take derivatives of Eqs. (\ref{eq_hjT}) and (\ref{eq_FT}), yielding
\begin{widetext}
\begin{align}
\label{eq_hjT}
\frac{d h_j^{(i+1)}}{dT} &= \frac{dh_j^{(i)}}{dT} + \ln\Bigg[\frac{e^{\frac{1}{T}(J_{ij} + h_i^{(i)})} + 1}{e^{\frac{1}{T}h_i^{(i)}} + 1} \Bigg] - \frac{1}{1 + e^{-\frac{1}{T}h_i^{(i)}}}\Bigg(\frac{d h^{(i)}_i}{dT} - \frac{h^{(i)}_i}{T}\Bigg) + \frac{1}{1 + e^{-\frac{1}{T}(J_{ij} + h_i^{(i)})}}\Bigg(\frac{d h^{(i)}_i}{dT} - \frac{J_{ij} + h^{(i)}_i}{T}\Bigg), \\
\label{eq_FT}
\frac{d F^{(i+1)}}{dT} &= \frac{dF^{(i)}}{dT} - \ln\big[e^{\frac{1}{T}h_i^{(i)}} + 1 \big] - \frac{1}{1 + e^{-\frac{1}{T}h_i^{(i)}}}\Bigg(\frac{d h^{(i)}_i}{dT} - \frac{h^{(i)}_i}{T}\Bigg).
\end{align}
\end{widetext}
Iterating the above equations, we arrive at the derivative $\frac{d\mathcal{F}}{dT} = \frac{d F^{(N+1)}}{dT}$, which completes our calculation of the average energy [Eq.~(\ref{eq_Eavg})].

The heat capacity is given by
\begin{equation}
C = \frac{d \langle E(\bm{x})\rangle}{dT} = -T \frac{d^2\mathcal{F}}{d T^2}.
\end{equation}
Taking derivatives of Eqs. (\ref{eq_hjT}) and (\ref{eq_FT}), we have
\begin{widetext}
\begin{align}
\label{eq_hjT2}
\frac{d^2 h_j^{(i+1)}}{dT^2} &= \frac{d^2 h_j^{(i)}}{dT^2} - \frac{1}{1 + e^{-\frac{1}{T}h_i^{(i)}}}\frac{d^2 h^{(i)}_i}{dT^2} - \frac{1}{T}\frac{e^{-\frac{1}{T}h^{(i)}_i}}{\big(1 + e^{-\frac{1}{T}h^{(i)}_i}\big)^2}\Bigg(\frac{d h^{(i)}_i}{dT} - \frac{h^{(i)}_i}{T}\Bigg)^2  \\
& \quad\quad\quad\quad\, + \frac{1}{1 + e^{-\frac{1}{T}(J_{ij} + h_i^{(i)})}}\frac{d^2 h^{(i)}_i}{dT^2} + \frac{1}{T}\frac{e^{-\frac{1}{T}(J_{ij} + h^{(i)}_i)}}{\big(1 + e^{-\frac{1}{T}(J_{ij} + h^{(i)}_i)}\big)^2}\Bigg(\frac{d h^{(i)}_i}{dT} - \frac{J_{ij} + h^{(i)}_i}{T}\Bigg)^2, \nonumber \\
\label{eq_FT2}
\frac{d^2 F^{(i+1)}}{dT^2} &= \frac{d^2 F^{(i)}}{dT^2} - \frac{1}{1 + e^{-\frac{1}{T}h_i^{(i)}}}\frac{d^2 h^{(i)}_i}{dT^2} - \frac{1}{T}\frac{e^{-\frac{1}{T}h^{(i)}_i}}{\big(1 + e^{-\frac{1}{T}h^{(i)}_i}\big)^2}\Bigg(\frac{d h^{(i)}_i}{dT} - \frac{h^{(i)}_i}{T}\Bigg)^2.
\end{align}
\end{widetext}
Finally, after computing $\frac{d^2\mathcal{F}}{d T^2} = \frac{d^2 F^{(N+1)}}{d T^2}$ iteratively, we have arrived at the heat capacity $C$.

\bibliography{MaxEntBib}

\end{document}